\documentclass[10pt,a4paper]{article}

\usepackage{graphicx}

\begin{document}
\begin{center}
\vspace*{12pt}
{\Large \bf Citations and gender diversity in reciprocal acknowledgement networks}
\vspace{12pt} \\
Keigo Kusumegi, Yukie Sano \\
\end{center}

\begin{abstract} % abstract
Acknowledgements in scientific articles suggest not only gratitude, but also the interactions among scientists. In this study, we examine the acknowledgement interactions employing data from open-access journals (PLOS series).
We built an acknowledgement network where the nodes represent authors and acknowledged people, while the links correspond to being mentioned in acknowledgements. Employing motif analysis, we showed how acknowledgement networks have developed, and how reciprocal relationships tend to emerge. To better understand these reciprocal relationships, we analysed the reciprocal sub-graphs of acknowledgement from two perspectives: citations and gender diversity. 
Firstly, we counted the number of citations, from both reciprocal and non-reciprocal authors. 
We found that reciprocal authors predominantly tend to cite other reciprocal authors rather than non-reciprocal ones. 
For gender diversity, we found that reciprocal pairs that include females tend to emerge more than male-male pairs of reciprocity in various fields, despite the fewer number of females. 
%Our results will shed new light on networks in academia.
\end{abstract}

\section{Introduction}
Science has developed through networks. This networks are often represented by the connections among scientists and papers, such as collaborations and citations. As a result, there are well-organized databases of this information, and numerous studies have analysed them. For example, collaboration networks, where nodes represent scientists, and two scientists are connected if they have collaborated, showed similar properties even when scientists come from different fields. The link distribution showed a fat-tailed distribution, where a few scientists have many collaborators, while most scientists have only a few collaborators~\cite{Newman2004}. 

In citation networks, where nodes represent papers or authors, and two nodes are connected if one has cited another; it is known that a community structure is created~\cite{Chen2010}, and highly cited authors tend not to collaborate but rather cite each other~\cite{Ding2011}. Furthermore, journal citation networks were also examined, where nodes represent journals.  These networks have high reciprocity and positive assortativity, which means that the nodes tend to connect to other nodes that have a similar degree~\cite{Franceschet2012}.  

Because the network of science is widely revealed from citation and collaboration relationships, it is still not enough to describe the development of science. For example, in the 1970s, one prominent paper on computer simulation in physics was published by three authors: Fermi, Pasta, and Ulam~\cite{Fermi1955}. 
As it turned out later, one more scientist named Tsingou, who actually wrote the computer simulation code, 
has not been recognized as importance contributor of the work~\cite{Dauxois2008}. 
Because the report states the following: ``Report written by Fermi, Pasta, and Ulam. Work done by Fermi, Pasta, Ulam, and Tsingou.'' 
Her name appeared on the acknowledgement statement as: ``We thank Miss Mary Tsingou for efficient coding of the problems and for runing the computations on the Los Alamos MANIAC machine.''~\cite{Fermi1955}

Additionally, it is common for a scientist to receive useful comments from other colleges and/or share biological facilities to promote their experiments. These academic activities are not listed as co-authors in published papers; they often appear in acknowledgement statements. 
Therefore, acknowledgement statements in published papers may suggest important contributors in the academic world.

Because acknowledgements in scientific papers are appreciation from authors to some entity, such as thanks for helpful advice or financial help, they could be interpreted in several meanings, for examples, ``scientific debt''~\cite{Edge1979} or ``sub-authorship collaboration''~\cite{Cronin2003}. 
In acknowledgements, the contribution types vary across the scientific fields, that is, sharing biological materials for biology and discussions for mathematics~\cite{Paul2017}. 
The distribution of acknowledged entities such as persons and institutions follows a power-law across various fields~\cite{Giles2004}. 
It was also said that acknowledgement has gradually become a constitutive element of academic writing, and so has the increase of collaboration in research~\cite{Cronin2004}. These studies were organised by extracting the necessary information from the acknowledgements without a network perspective. 
Acknowledgement network analysis has been performed mainly for funding interests~\cite{Wang2011, Mejia2018, Paul2016}. 
For instance, the Web of Science, one of the world's largest online academic databases provided by Clarivate Analytics, started collecting funding data described in the acknowledgement section for its publications, making it easier to track funding flow. 

However, little is known about acknowledgement networks at the level of human entities, and this is partly because of the difficulty of collecting data. 
Acknowledgements are not described in every published paper and do not have a defined format. Some works succeed in collecting acknowledged persons with high accuracy using machine learning techniques, such as SVM and regular expression process~\cite{Councill2005}. 
In this work, we employed data-driven analysis on acknowledgement networks to answer two main research questions: (1) how is the topology of research acknowledgement networks, and (2) what role play reciprocal relationships in acknowledgements from the point of view of citations and gender.

Understanding acknowledgement networks based on human interactions may provide new perspective of relationships in academia that can be captured in neither citation nor collaboration networks. To dig deeper into acknowledgement networks, we need first to understand their topological structure. Regarding the relationship between acknowledgements and citations, we focused on the significance of reciprocity in the context of acknowledgements, because reciprocity in human interaction has great importance in human behaviour and in social influence~\cite{Mahmoodi2018, Molm2007}. Furthermore, reciprocity is said to play a key role in innovation~\cite{Linton2000}; therefore, we examined the reciprocal relationships in acknowledgements from the viewpoint of gender diversity.

We first show basic information about acknowledgement networks. Next, we show the result of reciprocal acknowledgement and citation count relationships by simply comparing the statistical significance of citation counts between reciprocal authors and non-reciprocal authors. Moreover, we explored reciprocal acknowledgement networks from the perspective of the citation and gender diversity. Details regarding the methods and data are presented in the last section.

%\cite{koon,oreg,khar,zvai,xjon,schn,pond,smith,marg,hunn,advi,koha,mouse}

\section{Result}
\subsection{Overview of acknowledgement networks}
\begin{figure}[h!]
      \includegraphics[angle=90,width=1.2\linewidth]{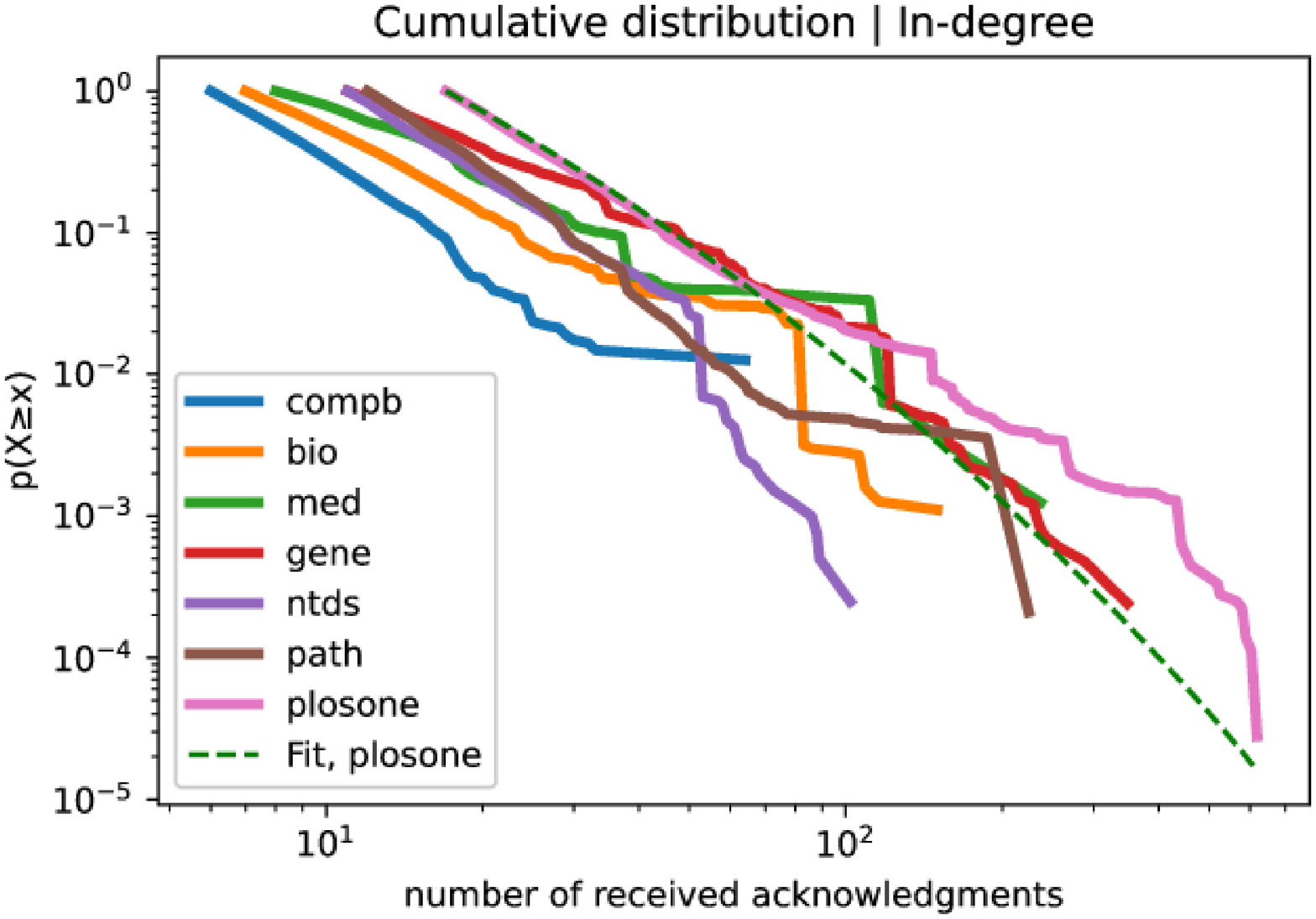}
  \caption{Cumulative in- and out- degree distribution of acknowledgements networks. The distribution follows a lognormal distribution where $ \mu = 4.1 \times 10^{-8}$, $\sigma= 1.1228$ and $x_{min}$ $(x>0): 11.0$ for {\tt PLOS ONE} (in-degree).}
    \label{Fig1}
      \end{figure}

In this work, we collected the description of acknowledgement statements from one of the biggest open access journals, the PLOS series (https://plos.org/).
We collected 214,645 papers published between 2006 and 2017 from seven major fields: {\tt biology}, {\tt computational biology}, {\tt genetics}, {\tt medicine}, {\tt NTDs (Neglected Tropical Diseases)}, {\tt pathogens}, and {\tt PLOS ONE} (interdisciplinary, main category). 
Each paper contains author names, published year and names mentioned in acknowledgement statement if they exit. 
Using these data, we created acknowledgement networks in each field where the nodes represent authors and people mentioned in acknowledgements, and the direct links connect authors to people mentioned in acknowledgements (see Methods for details).
Both, the in- and out-degree distributions, follow a lognormal function, where the logarithm of link $x$ is distributed normally as follows:
\begin{equation}
    f(x) = \frac{1}{\sqrt{2 \pi \sigma x}} \exp \left( -\frac{(\ln x -\mu)^2}{2 \sigma^2}\right)
\end{equation}
where $\mu$ and $\sigma$ are the mean and the standard deviation of the normal distribution, which have different values depending on the field (Figure~\ref{Fig1}). The highest $\mu$ is 6.68 in {\tt medicine}, 
while the lowest is 1.83 in {\tt computational biology}. This lognormal distribution suggests that most of the authors mention a small number of people in their acknowledgements, while a few of them mention a huge number. 

We also calculated the global clustering coefficient $C$, where triangle is a set of three nodes, each of whom is connected to the other two nodes. $C = 1$ for a fully connected graph, and $C = 0$ for a random graph without triangles. The $C$ for all datasets was small ($C < 0.02$) (Table~\ref{Tab1}). 

\subsection{Motif analysis}
To clarify local connectivity in the acknowledgement network, we employed 3-node motif analysis over the largest connected component of each network. Motif analysis enables us to understand the pattern of connected nodes in networks from a microscale. In this analysis, all three connected nodes are assigned to one of the thirteen types of subgraphs, and those counts were normalised $z$-scores to be comparable with the others. 

\begin{figure}[h!]
      \includegraphics[angle=90,width=1.2\linewidth]{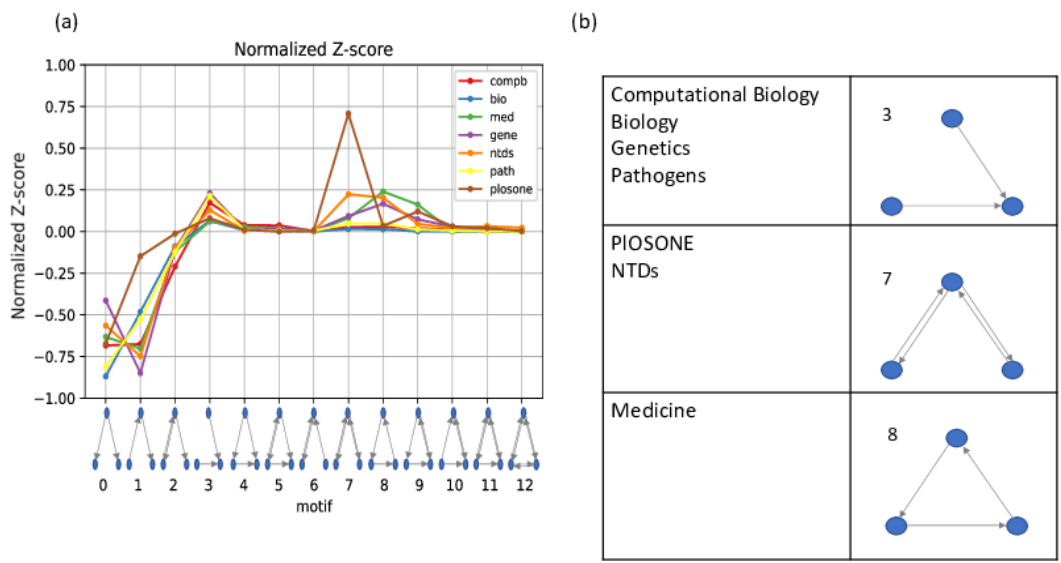}
  \caption{Motifs in acknowledgement networks. (a) Normalised $Z$-score over 13 motifs. (b) The most frequently appeared motifs in each field.}
  \label{Fig2}
\end{figure}

Some networks show similar frequent patterns from the $z$-score (Figure~\ref{Fig2}(a)). Motifs 3 and 7 tend to appear in all fields except for {\tt medicine}. This coincides with the low clustering coefficient, because triangle patterns do not appear frequently. In addition, $z$-scores of most motifs, including reciprocity (e.g., motif 7 and 9) were positive over all datasets. This suggests that reciprocity tends to be created in the acknowledgement network. Therefore, we supposed that reciprocity in the acknowledgement network might imply some strong and special ties between the two authors. For instance, reciprocity might be a sign of joining collaboration in the future.

\subsection{Reciprocal relationships}
\begin{figure}[h!]
      \includegraphics[angle=90,width=1.2\linewidth]{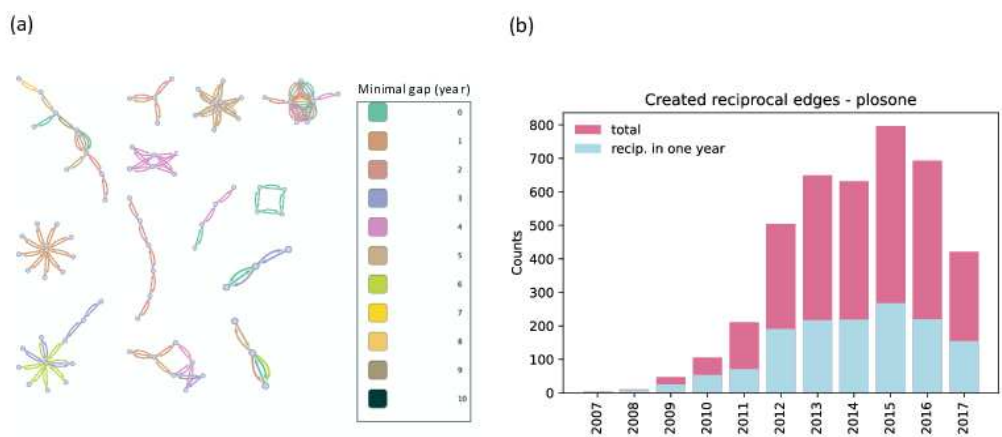}
  \caption{Time difference to complete reciprocal links.
       (a) An example of reciprocal subgraphs of acknowledgement network. Link colour represents the minimal gap year of reciprocal acknowledgements within a reciprocal pair. 
(b) The number of created links by year. The red bar shows the total number of reciprocities in a year, while sky-blue shows the number of reciprocities in which both of them are sent in the same year.}
    \label{Fig3}
\end{figure}

Reciprocity has been investigated in social and economic sciences~\cite{Lewis2015, Thomson2006} and it is also important for collaboration. According to the citation relationship, the number of reciprocities is increasing, and reciprocal citations negatively correlate with a long-term successful academic career~\cite{Li2019}. Considering those works related to reciprocity, we show the characteristics of reciprocity in acknowledgement networks.
Figure~\ref{Fig3}(a) illustrates parts of reciprocal subgraphs, where the link colour implies the gap time. The gap time is the minimum difference of the mentioned year of any two reciprocal acknowledgements. For example, the green colour in Fig.~\ref{Fig3}(a) means that the gap time is 0, that is, the reciprocal acknowledgements correspond to the same year. Figure~\ref{Fig3}(a) also shows that it takes a few years to complete reciprocal links.

In Fig.~\ref{Fig3}(b), purple bars show the total number of reciprocities and the sky-blue bars indicate the number of reciprocities in which both acknowledgements had been written in the same year. There is an increasing number of reciprocities after 2009, and a certain portion of these is completed within the same year. The average gap time between reciprocities is 1.48 years; they seem to be completed in a relatively short term.

\subsection{Reciprocity and citation counts}
Citation count is often used as an index of the scientist's profile. To examine the impact of reciprocal acknowledgement relationships on citation, we separated nodes in an acknowledgement network into reciprocal and non-reciprocal nodes. 
The reciprocal nodes contain only pairs of authors who have a reciprocal relationship, such as illustrated by subgraphs in Fig.~\ref{Fig3}(a), and non-reciprocal nodes contain the rest of the subgraphs. 

\begin{figure}[h!]
      \includegraphics[angle=90,width=1.5\linewidth]{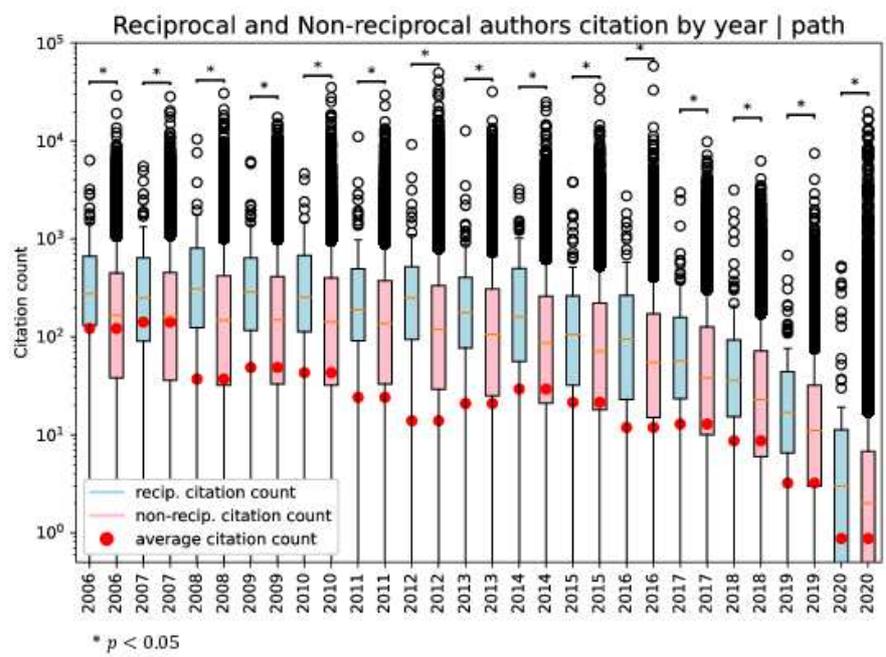}
  \caption{Citation counts of reciprocal and non-reciprocal authors by year in {\tt pathogen}.
      Sky-blue shows the citation of reciprocal authors and pink shows that of non-reciprocal authors. 
Red dot shows the average number of citations counts in these fields.}
    \label{Fig4}
\end{figure}

Citation data was collected from Microsoft Academic Search API~
\cite{Tang2008, Sinha2015}, which allows us to collect data from Microsoft Academic Graph (MAG). 
The data includes the author's name, the citation count, the paper ID and referenced papers. 
Because the average number of citations varies by research field, we averaged other scientists' citations in the same research field as a baseline (red dots in Fig.~\ref{Fig4}). The number of extracted citation data is depicted in the Methods section.

Figure~\ref{Fig4} shows the authors' annual citation counts in {\tt pathogens} grouped by reciprocity and non-reciprocity authors (the results of the other fields are shown in the Appendix). Using the Mann-Whitney U test, we found that reciprocal authors are statistically cited more than non-reciprocal authors in most of the given years. 
{\tt pathogens}, {\tt computational biology}, {\tt NTDs}, and {\tt PLOS ONE} showed similar results. Regarding this analysis, we excluded {\tt  medicine} and {\tt genetics} because the amount of citation data was too small for an adequate Mann-Whitney U test. Although the average citation of reciprocal and non-reciprocal authors from our dataset was slightly higher than the baseline from the MAG dataset, the difference between reciprocal and non-reciprocal authors was still present.

\subsection{Reciprocal authors citing from other reciprocal authors}
As we showed in the previous section, reciprocal authors tend to be cited more frequently than non-reciprocal authors. To reveal the mechanism, we posited that reciprocal authors might cite the paper of other reciprocal authors, that is, there might be a direct overlap between the acknowledgement network and the citation network. 
To answer this question, we compared the number of reciprocal authors cited from other reciprocal authors and those of non-reciprocal authors cited from reciprocal authors. 

We tested whether there was a significant difference in the number of reciprocal authors' papers in the list of referenced articles between those two groups using the Mann-Whitney U test. 
As a result, we found that reciprocal authors statistically cite the other reciprocal authors more than non-reciprocal authors (Table~\ref{Tab2}). The average number of reciprocal authors' papers cited by other reciprocal authors is more than twice the number of those cited by non-reciprocal authors. 

Meanwhile, the percentage of citation counts between two authors who are under the same reciprocal acknowledgement relationship is low, from 0.43\% to 13.27\% of total citations within reciprocal groups. Thus, acknowledgement reciprocity relationships do not directly correspond with citation relationships. 
This suggests that reciprocal authors belong a large community where scientists are citing and mentioning acknowledgement each other.

\subsection{Reciprocity and gender diversity}
Gender diversity is said to be one of the key elements for success~\cite{Nielsen2017}. In addition to the citation relationship, we examined gender diversity in reciprocal pairs. 
We assumed that gender-related topology would reflect the reciprocal group's citation, like author ships and citations~\cite{West2013, Maliniak2013, Dworkin2020}. To identify the author's gender, we applied the Gender API (https://gender-api.com/), which is a gender-telling service using over three million validated names from 191 countries. This database detects accurately whether a name belongs to a male or a female. In our data, 9 in 10 reciprocal authors' gender are detected with more than 80\% accuracy over all datasets. In addition, 500 non-reciprocal authors' gender in each field dataset are used as a comparison with reciprocal authors.

\begin{figure}[h!]
 \includegraphics[angle=90,width=1.2\linewidth]{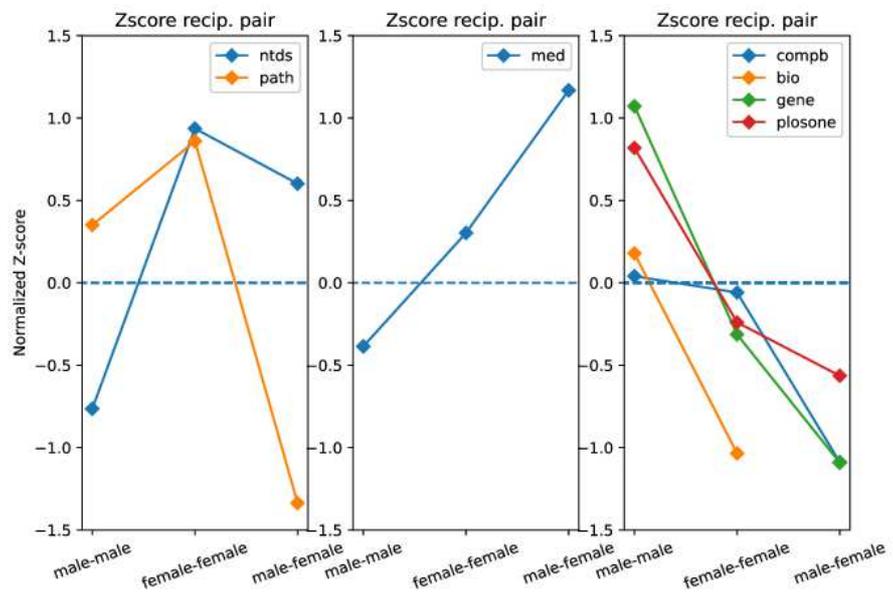}
  \caption{Normalised $Z$-score of 3 types of pairs, which is male-male, female-female and male-female. By the pattern of the highest $Z$-scores, the seven fields are divided into three groups.}
    \label{Fig5}
\end{figure}

Because the majority of the nodes in the network correspond to male scientists, 53\% in {\tt medicine} and 75\% in {\tt computational biology}, we computed normalised $z$-scores of three different pairs of reciprocity: male–male, female–female, and male–female. Figure~\ref{Fig5} shows the normalised $z$-scores of reciprocal pairs. First, male–male pairs are highly likely to appear in more than half of the research fields such as {\tt PLOS ONE}. However, male–female or female–female are more likely to show in three research fields despite the smaller number of females. For example, the male-male pair is less likely to happen in {\tt NTDs} and {\tt medicine}. This suggests that a pair of gender in reciprocity is heterogeneous, and it depends on the research fields. 
%Although gender diversity cannot fully support the result that reciprocal authors tend to get more citations, it is key for studying the acknowledgement network science and it remains as a future task. 

\section{Discussion}
Acknowledgement statements in published papers have various roles, such as representing gratitude or strategically avoiding potential referees. 
Here we explored networks in academia from the perspective of acknowledgements. Here, we built a network by representing authors or people mentioned in acknowledgement as nodes, and the links correspond to being mentioned in acknowledgements. Basic network structures such as degree distributions, clustering coefficients and motifs reveal topological information of acknowledgement networks from microscale to macroscale. This reveals that the topology of acknowledgement networks differs in each research field. From motif analysis, reciprocal relationships tend to emerge in acknowledgement networks. 

To uncover the creating mechanism of reciprocal relationships in acknowledgement, we explored the citation relationships and gender diversity. In the citation relationship, we showed that reciprocal authors tend to cite other reciprocal authors more frequently than non-reciprocal authors. Although reciprocal authors have more strained citation relationships among them, non-reciprocal authors also cite the papers of reciprocal authors to some extent. This fact could explain why reciprocal authors might be well-known authors in their research field, and consequently, they made strong citation and acknowledgement relationships with same level authors. 

From the angle of gender diversity, since the male–male pair is dominant in most research fields, including {\tt PLOS ONE}, the female–female pair emerged more in the fields of {\tt biology}, {\tt pathogens}, and {\tt NTDs}, although the number of females is less than the number of males in our dataset. It should be noted that we showed just divergence of a pair of reciprocal pairs, and it could not directly explain the effect of the citation number of reciprocal authors. We anticipate that the gender diversity and acknowledgement networks perspectives will contribute to the science of success in the future.

There remains the bias of datasets that are only based on PLOS series as a limitation. Because of the data collection difficulty, it takes time to collect an amount of acknowledgement data, we collected only fully named entities to avoid the confusion. This makes acknowledgement network analysis hard to capture the entire scientist's world as much as citation and collaboration network analysis do. When a database of acknowledgments might be developed in the future, acknowledgment networks would be analysed more precisely.
Also, building a multilayer network~\cite{Kivela2014} where each layer represents collaboration, citation and acknowledgement network is an interesting direction for further exploration.

\section{Methods}
\subsection{Acknowledgement data}
We collected text data from PLOS (https://plos.org/), which has published 214,645 papers from 2006 to 2017. PLOS was officially launched in 2006 and had become, by 2010, the biggest open-access journal. Because PLOS publishes multidisciplinary subjects, it covers a variety of fields, from medicine to the humanities. We collected 214,645 papers from seven research fields: {\tt computational biology}, {\tt biology}, {\tt medicine}, {\tt genetics}, {\tt pathogens}, {\tt NTDs} and {\tt PLOS ONE} (Table~\ref{Tab1}).

From the collected data, we extracted basic information: author's name and year of the publication. We also extracted the acknowledgement statements as raw texts. 
We then employed Stanford CoreNLP (https://corenlp.run/) to phrase and identify the mentioned authors in the acknowledgement texts. 
In the present study, only fully named entities were used as available mentioned names in order to avoid confusion among people with the same initial expressions. Note that we removed the names of ``Albert Einstein'' and ``Mary Curie'', which appeared as institution's name.
Among the 214,645 papers collected, 71\% contained acknowledgements and 338,027 names were mentioned in it. 
In {\tt PLOS ONE}, 130,774 papers had acknowledgement statements and 277,545 names were mentioned in the acknowledgements. In {\tt medicine}, 52\% of papers include acknowledgements, while in {\tt genetics} and {\tt pathogens} over 80\% of papers contain acknowledgements. 
The average number of names mentioned per paper is between two and three in {\tt biology}, {\tt genetics}, {\tt NTDs}, {\tt pathogens}, and {\tt PLOS ONE}. In {\tt computational biology}, on average less than two scientists were mentioned, while an average of 3.5 scientists were mentioned in {\tt medicine}. The average number of co-authors per paper varies widely from 4.3 in {\tt computational biology} to 9.2 in {\tt genetics} (Table~\ref{Tab1}).

\subsection{Acknowledgement network}
The acknowledgement network is defined as a weighted directed graph $G [V, E, W]$, where $V$ denotes a set of nodes, $E \subseteq\{(s, t) | s, t \in V~\mathrm{and}~s \neq t\}$, denotes a set of links and $W$ denotes the weight of the links, which depends on the number of co-authors. 
The nodes represent names who wrote acknowledgements or were mentioned in acknowledgements. 
The directed links from $s$ to $t$ indicate pairs of nodes $(s, t)$, where node $s$ acknowledges node $t$.  In a paper written by $n$ authors, we assumed that the acknowledgement came for all authors and assigned a weight of $1/n$ for each author. 

\subsection{Motif analysis}
Statistical significance of a motif $M$ is defined by the $z$-score $Z_M$:
\begin{equation}
    Z_M=\frac{N_M - \langle N^{rand}_{M} \rangle}{\sigma^{rand}_{M}},
\end{equation}
where $N_M$ is the number of motif $M$ in the given network, 
and $\langle N^{rand}_{M} \rangle$  and $\sigma^{rand}_{M}$ are the mean and standard deviation of $M$'s occurrence in the set of random networks, respectively. 
Random networks correspond to the configuration model in which the pairs of links are stochastically exchanged, while the number of nodes and links are maintained. Then, $Z_M$ is normalised as follows:
\begin{equation}
    Z_M^{norm} = \frac{Z_M}{\sqrt{\Sigma{Z_M^2}}}
\end{equation}

\subsection{Reciprocal and non-reciprocal authors}
The acknowledgement network was divided into two subgraphs: reciprocal and non-reciprocal. The reciprocal subgraph was composed of the nodes that are connected by reciprocal acknowledgements, while the non-reciprocal subgraph contains all the other nodes. Then, we extracted nodes (authors) whose citation data were available from any of the subgraphs.

\subsection{Reciprocity - Citation}
\begin{figure}[h!]
  \includegraphics[angle=90,width=1.2\linewidth]{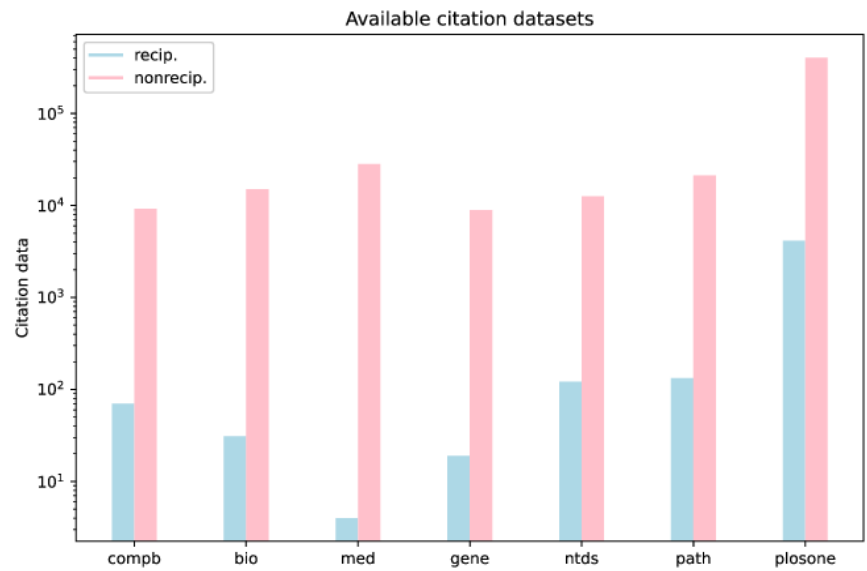}
  \caption{Extracted citation data from Microsoft Academic Search API. Sky-blue bars show the number of papers authors by reciprocal authors and pink bars show by non-reciprocal authors. }
    \label{Fig6}
\end{figure}

Citation data were extracted from Microsoft Academic Graph (MAG) using Microsoft Academic Search API querying the author's names. The data contained authors' IDs, papers' IDs, the list of referenced paper IDs and citation counts for each paper. We collected all available 504,796 authors' citation information. Figure~\ref{Fig6} shows the number of available citation data by reciprocity and research field. Regarding citation-related analysis, the datasets of {\tt medicine} and {\tt genetics} were excluded because the data of reciprocity were not sufficient to perform a Mann-Whitney U test. 
We computed two types of citations: citations where a reciprocal author cites the paper of another reciprocal author and citations where a non-reciprocal author cites the paper of a reciprocal author. Next, we compared the differences in citation counts between these two types of citations using the Mann-Whitney U test. Moreover, we calculated the number of citations between reciprocal authors.

\subsection{Reciprocity - Gender}
To verify which pair of gender on reciprocity appears frequently, we computed the statistical significance $z$-score as we did in motif analysis. Here, $M$ is a set of male–male, female–female, and male–female. 
$\langle N_M^{rand} \rangle$and $\sigma_M^{rand}$ are the mean and standard deviation of $M$'s occurrence in the set of random networks, respectively. 
In random networks, the node is assigned to male or female with the probability of the fraction of the number of males or females in the network.

%%%%%%%%%%%%%%%%%%%%%%%%%%%%%%%%%%%
%%                               %%
%% Figures                       %%
%%                               %%
%% NB: this is for captions and  %%
%% Titles. All graphics must be  %%
%% submitted separately and NOT  %%
%% included in the Tex document  %%
%%                               %%
%%%%%%%%%%%%%%%%%%%%%%%%%%%%%%%%%%%

%%
%% Do not use \listoffigures as most will included as separate files
\clearpage
\section*{Figures}

%%%%%%%%%%%%%%%%%%%%%%%%%%%%%%%%%%%
%%                               %%
%% Tables                        %%
%%                               %%
%%%%%%%%%%%%%%%%%%%%%%%%%%%%%%%%%%%
\clearpage
%% Use of \listoftables is discouraged.
%%
\section*{Tables}

\begin{table}[h!]

\caption{Basic properties of acknowledgement network.}
\rotatebox{90}{
      \begin{tabular}{lccccccc}
        \hline
           & {\tt medicine}  & {\tt biology}   & {\tt comp.bio.} & {\tt NTDs} & {\tt pathogens} & {\tt genetics} & {\tt PLOS ONE}\\ \hline
        Number of papers & 3301 & 4372 & 5120 & 5256 & 6030 & 6638 & 183928\\
        The average number of co-authors & 6.5 & 5.5 & 4.3 & 8.1 & 7.8 & 9.2 & 6.8\\
        Proportion of paper \\contain acknowledgement & 0.523 & 0.672 & 0.7 & 0.756 & 0.824 & 0.865 & 0.711 \\
        Average number of acknowledged \\persons per paper & 3.54 & 3.03 & 1.82 & 2.35 & 2.55 & 2.95 & 2.12 \\
        Average in-degree & 6.69 & 2.37 & 1.84 & 1.98 & 1.93 & 3.06 & 2.26\\
        Average gap time \\for creating reciprocity [Years] & 0.91 & 3.51 & 2 & 0.98 & 1.92 & 2.54 & 1.49\\
        Average clustering coefficient $C$ & 0.005 & 0.003 & 0.012 & 0.012 & 0.008 & 0.008 & 0.013\\
        \hline
    \end{tabular}
    }
    \label{Tab1}  
\end{table}

\begin{table}[h!]
\caption{Comparison of reciprocal and non-reciprocal authors' citations.}
\rotatebox{90}{
      \begin{tabular}{lccccccc}
        \hline
           & {\tt medicine}  & {\tt biology}   & {\tt comp.bio.} & {\tt NTDs} & {\tt pathogens} & {\tt genetics} & {\tt PLOS ONE}\\ \hline
Number of citations \\by reciprocal authors & 4 & 31 & 70 & 121 & 133 & 19 & 4127 \\
Number of citations \\by non-reciprocal authors & 28201 & 15006 & 9229 & 12526 & 21265 & 8894 & 405170\\
Average number of reciprocal author's \\citation of other reciprocal authors  & 1 & 21.3 & 16.2 & 38.2 & 28.5 & 77.7 & 53.4\\
Average number of non-reciprocal author's \\citation of reciprocal authors  & 11.7 & 4.8 & 7.6 & 18.5 & 13.4 & 48.3 & 31.4\\
Ratio of citation between \\a pair of reciprocity against \\the number of citations \\within reciprocal authors & 0.0043 & 0.0748 & 0.0599 & 0.1275 & 0.1327 & 0.1218 & 0.0637\\
        \hline
    \end{tabular}
    }
    \label{Tab2}  
\end{table}

%%%%%%%%%%%%%%%%%%%%%%%%%%%%%%%%%%%
%%                               %%
%% Additional Files              %%
%%                               %%
%%%%%%%%%%%%%%%%%%%%%%%%%%%%%%%%%%%

%\section*{Additional Files}
%  \subsection*{Additional file 1 --- Sample additional file title}
  
%    Additional file descriptions text (including details of how to
%    view the file, if it is in a non-standard format or the file extension).  This might
%    refer to a multi-page table or a figure.

%  \subsection*{Additional file 2 --- Sample additional file title}
%    Additional file descriptions text.

%\end{backmatter}
\end{document}